\documentclass{spie}

\usepackage{amsmath,amsfonts,amssymb}
\usepackage{graphicx}
\usepackage[colorlinks=true, allcolors=blue]{hyperref}
\usepackage{float}
\usepackage{array}
\newcolumntype{P}[1]{>{\centering\arraybackslash}p{#1}} 
\usepackage{xcolor}
\usepackage{gensymb}
\usepackage{textcomp}
\usepackage{soul}
\title{Faster imaging simulation through complex systems: a coronagraphic example}

\author[a]{Kian Milani}
\affil[a]{James C. Wyant College of Optical Sciences, University of Arizona}
\author[b]{Ewan S. Douglas}
\affil[b]{Steward Observatory, University of Arizona \\
Massachusetts Institute of Technology}

\begin{document} 
\maketitle

\begin{abstract}
    End-to-end simulation of the influence of the optical train on the observed scene is important across optics and is particularly important for predicting the science yield of astronomical telescopes. As a consequence of their goal of suppressing starlight, coronagraphic instruments for high-contrast imaging have particularly complex field-dependent point-spread-functions (PSFs). The Roman Coronagraph Instrument (CGI), Hybrid Lyot Coronagraph (HLC) is one example. The purpose of the HLC is to image exoplanets and exozodiacal dust in order to understand dynamics of solar systems. This paper details how images of exoplanets and exozodiacal dust are simulated using some of the most recent PSFs generated for the CGI HLC imaging mode. First, PSFs are generated using physical optics propagation techniques. Then, the angular offset of pixels in image scenes, such as exozodiacal dust models, are used to create a library of interpolated PSFs using interpolation and rotation techniques, such that the interpolated PSFs correspond to angular offsets of the pixels. This means interpolation needs only be done once and an image can then be simulated by multiplying the vector array of the model astrophysical scene by the matrix array of the interpolated PSF data. This substantially reduces the time required to generate image simulations by reducing the process to matrix multiplication, allowing for faster scene analysis. We will detail the steps required to generate coronagraphic scenes, quantify the speed-up of our matrix approach versus other implementations, and provide example code for users who wish to simulate their own scenes using publicly available HLC PSFs.
\end{abstract}

\section{Introduction} \label{sec:intro}
\begin{figure}[h]
    \centering
    \includegraphics[scale=0.5]{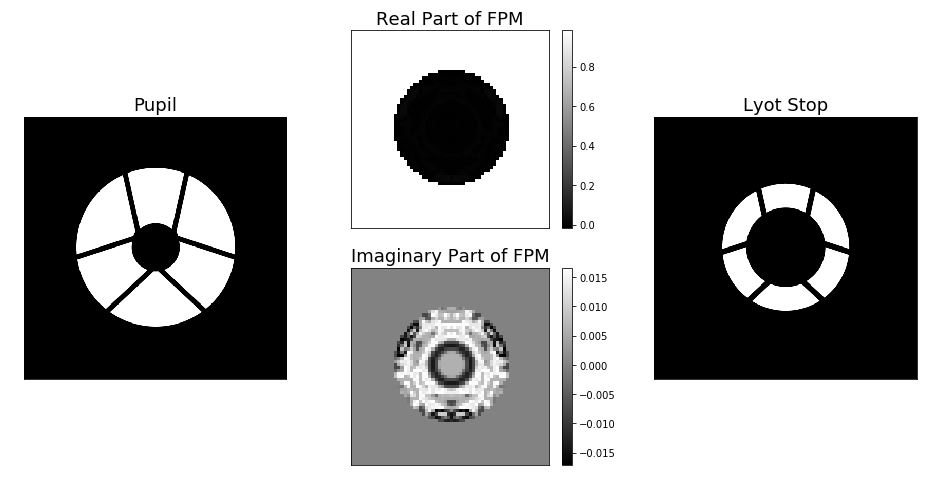}
    \caption{Important components within the deign of HLC as of Phase-B project status. The real and imaginary components of the FPM shown are for the central wavelength of 575nm. The pupil mask nor field stop are shown as those are standard circular apertures.}
    \label{fig:hlc_configuration}
\end{figure}

\begin{figure}[h]
    \centering
    \includegraphics[scale=0.55]{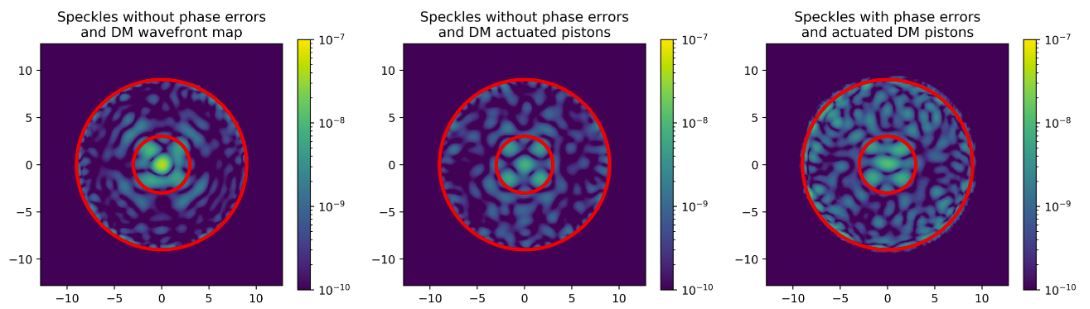}
    \caption{PROPER simulation of speckles within the coronagraphic field of the HLC in the image plane. The extent of the images are in $\lambda/D$ and the red circles define the inner and outer working angles respectively.}
    \label{fig:hlc_speckles}
\end{figure}

The Nancy Grace Roman Space Telescope is set to have two imaging instruments, one being the Wide Field Instrument (WFI) and the other being the Coronagraph Instrument (CGI). The CGI will be used for the search and study of exoplanets and exozodiacal dust close to individual stars. The CGI will have two modes of operation, one being the Shaped Pupil Coronagraph (SPC) and the other being the Hybrid Lyot Coronagraph (HLC) \cite{trauger_hybrid_2016}. The imaging configuration of interest in this paper is the CGI HLC. Figure \ref{fig:hlc_configuration} shows some of the main components of the HLC, which consists of a standard circular aperture, a complex focal plane mask (FPM) to affect both amplitude and phase, a Lyot stop, a field stop, a 575nm 10$\%$ bandpass filter, and an imager. The HLC will have an annular field of view (FOV) with inner and outer working angles of approximately 3$\lambda/D$ and 9$\lambda/D$ for which two deformable mirrors will be used to dig dark holes to further suppress the starlight that leaks into the image plane and improve the contrast. This makes the HLC suited for exoplanet searches and the study of exozodiacal dust\cite{wfirst2019}. Figure \ref{fig:hlc_speckles} shows the speckles within the coronagraphic field with and without optic phase errors. 

Throughout this paper, the simulated data was generated using PROPER models \cite{krist_proper:_2007,krist_numerical_2015,krist_wfirst_2018} of the HLC \footnote{Python Version 3.2.3, https://sourceforge.net/projects/proper-library/files/} along with PSF data files shared by IPAC\footnote{roman.ipac.caltech.edu}. The PSF models shared by the project team provide high-fidelity models of exoplanet PSFs. For modeling debris disks and exozodiacal dust, the computing time to generate a PSF for each field-point in the extended source is typically quite time consuming. This typically leads to simplifications, such as convolving the debris disk scene with a single off-axis PSF. An example of this is shown further in Section \ref{sec:simulation}, where a simulation done with convolution is compared to a simulation done with PSFs for every field-point of the scene. This work explores alternate approaches to speed up the computation of crowded or dusty astrophysical scenes while preserving the field-dependent properties of a coronagraphic image which are important to accurately modeling morphology and flux from disks close to the coronagraph inner working angle \cite{douglas_simulating_2019}. 

The process of PSF simulation starts with modeling the system in diffraction software. Once the model is created, propagation techniques can be applied to wavefronts in order to obtain the coronagraphic fields and PSFs. The most commonly used propagation techniques are Fraunhofer, Fresnel, and angular spectrum. While Fraunhofer is the fastest to compute, it is the least accurate given its plane-wave approximation, which is why the PSFs used for simulation in this paper are generated by Fresnel or angular spectrum methods. While these methods are not as computationally expensive as Rayleigh-Sommerfield diffraction algorithms, the results demonstrated by Krist\cite{kristTDEM} show agreeable data between the methods, meaning there is no great loss of accuracy with the chosen methods. 

The various off-axis PSFs are then used for interpolation across the angular offsets of the pixels within the given array of the scene to be simulated. The creation of the PSFs can be very computationally expensive due the exponential functions and two-dimensional Fourier Transforms that must be calculated for each propagation step\cite{douglas_accelerated_2018-1}, therefore, it is unreasonable to use propagation to calculate PSFs for every field-point within the extended source to be simulated as will be shown in Section \ref{sec:results_creation} with the timing results of the PSF creation. To avoid such a procedure, a set of PSFs along a single axis in the image can be interpolated and rotated to act as the PSFs of various field points in the extended source. While interpolation and rotation of the PSFs is not as time consuming as the propagation of wavefronts, if the process is repeated each time a simulation is run instead of using previously computed interpolated PSFs, significantly more time would be required. Two methods of interpolation over 2D grids are tested and compared, which are nearest neighbor and linear interpolation. 

\begin{figure}[h]
    \centering
    \includegraphics[scale=0.3]{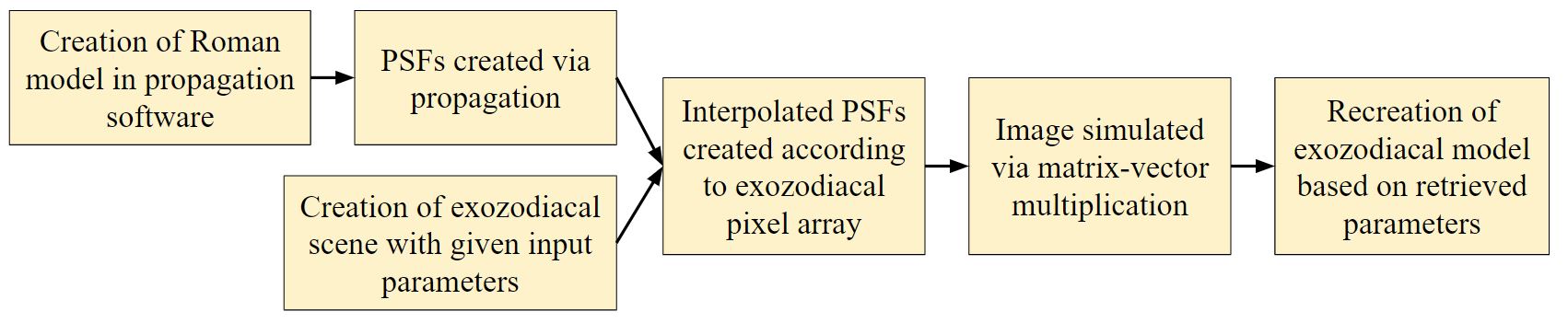}
    \caption{Flow chart which conveys how the simulations are created and analyzed to extract zodiacal parameters back from the simulations. The significance of having faster simulations in this application is greater due to the fact the MCMC algorithm would require far more time to converge, although it would do so in the same number of steps.}
    \label{fig:roman_chart}
\end{figure}

Figure \ref{fig:roman_chart} displays the sequence of steps required to generate the simulations of the exozodiacal scenes. Following the creation of the interpolated PSF arrays, a linear optical system is assumed, allowing for simulations to be performed by matrix-vector multiplication. In order to perform the matrix-vector multiplication, one may not be able to load the entire matrix into computer memory given the size of the arrays, however, individual rows or chunks can be loaded in using loops to manually perform the multiplication.


This paper will sequentially cover how to use interpolation to create arrays of interpolated PSF data, how to generate an image simulation of the astrophysical scene using a single matrix-vector multiplication, and the results from the simulations. Within the results, the time required to generate image simulations are compared along with the numerical differences between the simulations themselves due to the fact that different interpolation methods are used across the 2D grids. The code used to generate the data and figures can be found on Zenodo\cite{kian1393_2020_3965084}\footnote{https://zenodo.org/record/3965084}. 

\section{Creating the Interpolated PSF Library} \label{sec:create_psfs}
Creation of the interpolated PSFs begins by setting up an interpolating function with the use of a grid interpolator. As mentioned above, only nearest or linear interpolations are employed with the use of the RegularGridInterpolator() provided by the Scipy module. The variables for this function are a datacube of PSFs with angular offsets along one dimension of the image, which were created using the propagation techniques mentioned above, along with the information regarding what the offsets and array dimensions of the PSFs. Therefore, when the function is given an offset, it will return an array that is the interpolated PSF for the given offset. Given the datacube of PSFs used for interpolation only contains offsets along one axis of the image, this interpolating function alone cannot be used to interpolate for PSFs with offsets in both the $x$ and $y$ directions. In order to do so without using propagation to create a larger datacube of propagated PSFs, the rotate() function, also found within the Scipy module, is employed to rotate the PSF to the correct position of the offset. 

One of the sets of the PSFs used in this paper for interpolation are the Roman CGI Off-axis PSFs for the HLC imaging mode provided by Hanying Zhou and are publicly available on the \href{https://roman.ipac.caltech.edu/sims/off_axis_PSF.html}{Roman website}. There are two sets of PSFs, one of which has offsets only in the $x$-direction and another that has offsets in both the $x$ and $y$ directions. As mentioned above, only the 1D offset PSFs are what were chosen to be used for this paper. In this datacube, there are 85 PSFs, where the offset sampling for the PSFs is 0.06$\lambda/D$ within $4\lambda/D$, 0.5$\lambda/D$ within $4-9\lambda/D$, and 0.25$\lambda/D$ within $9-11\lambda/D$. These offsets are then converted to units of milliarcseconds for ease of use. It is notable that the reason for the unequispaced sampling for the PSF offsets is due to the evolution of the PSF near the inner and outer working angle of the HLC, although the evolution is not as significant near the outer working angle. In section \ref{sec:results_sims}, analysis is done to compare whether the low sampling within the dark hole region of the HLC has substantial affects on the image simulations of the scenes by also using PSFs with high sampling all throughout the HLC FOV. Examples of the publicly available PSFs are shown below in Figure \ref{fig:psf_examples}.

\begin{figure}[h]
    \centering
    \includegraphics[scale=0.4]{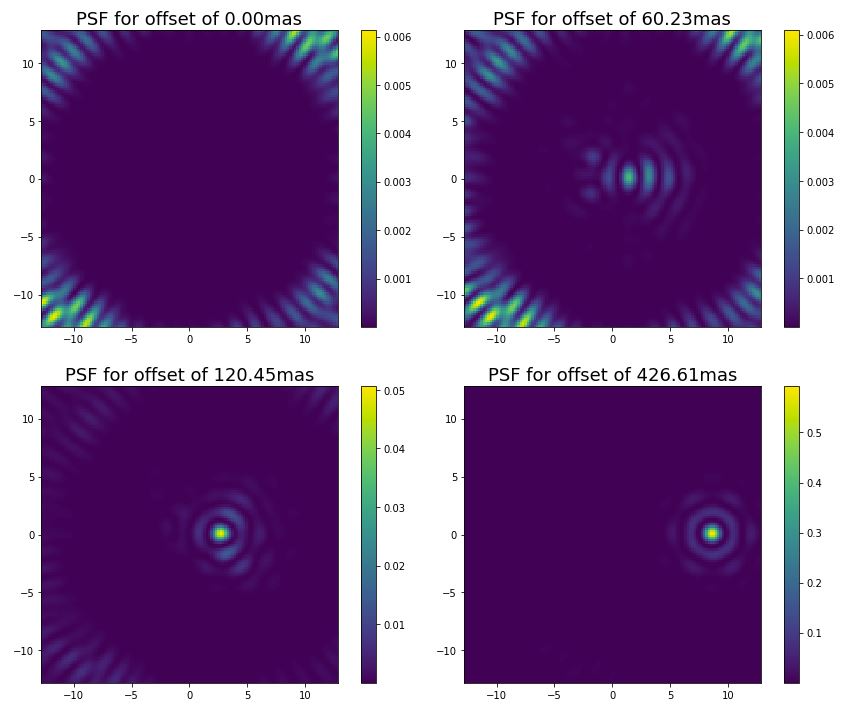}
    \caption{PSFs publicly available from IPAC with various source offsets in milliarcseconds. The extent of the images are given in units of $\lambda/D$.}
    \label{fig:psf_examples}
\end{figure}

\begin{figure}[h]
    \centering
    \includegraphics[scale=0.3]{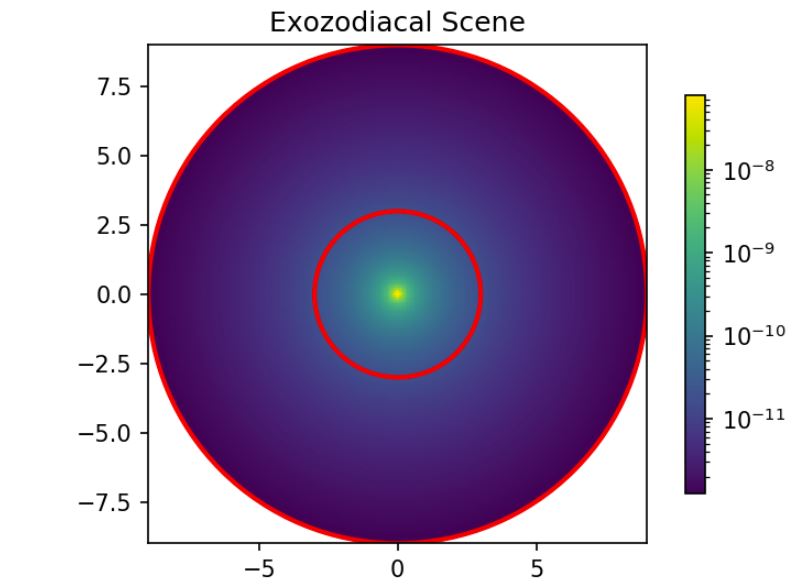}
    \caption{Example of a 256x256px zodiacal scene with a pixel scale of 3.529mas, resulting in the extent of the axis to be defined from $-9\lambda/D$ to $9\lambda/D$. The units of the pixels representing flux in the scene are Janskys [Jy]. The inclination of this dust model is 0\degree, the distance to the scene is 10pc, and the radial power law of dust distribution is $\alpha=1.34$. The scene has been masked such that pixels outside of the outer working angle do not contribute to the flux in the image.}
    \label{fig:zodi_example}
\end{figure}

Aside from the offset information for the PSFs, the parameters needed to fully define the PSFs are the central wavelength of 575 nm, the pupil diameter of 2.3631m, and the pixel scale (plate scale) of the detector. However, to create the interpolated library, information regarding the scenes to be simulated is required as well. The zodiacal scenes used in this paper were 256x256px arrays with a pixel scale of 3.529mas generated with the publicly available IDL program ZODIPIC \cite{kuchner_zodipic_2012}, which was created by \href{https://asd.gsfc.nasa.gov/Marc.Kuchner/home.html}{Marc Kuckner}. A smooth zodiacal dust distribution was modeled with ZODIPIC defaults of a Sunlike star with zero rings, bands or blobs at a distance of 10 parsecs and an $\alpha=1.34$ for the radial power law of dust distribution. The pixel scale results in the 256x256px scenes to extend to the outer working angle of $9\lambda/D$. These parameters and their respective variables are defined below. Figure \ref{fig:zodi_example} displays an example of the zodiacal scenes used where the zodiacal dust inclination is 0\degree. 

\begin{equation}
    \begin{aligned}
        &\text{Central Wavelength:}\ \lambda = 575 \text{nm} \\
        &\text{Diameter of the telescope:}\ D = 2.3631 \text{m} \\
        &\text{Number of pixels in PSF arrays:}\ n_{px} = 128 \\
        &\text{Pixel scale of PSF arrays:}\ s_{px} = 0.2\lambda/D = 10.038\text{mas} \\
        &\text{Number of pixels in zodiacal model arrays:}\ N_{px} = 256 \\
        &\text{Pixel scale of zodiacal model arrays:}\ S_{px} = 3.529\text{mas} \\
    \end{aligned}
\end{equation}

In order to generate the interpolated PSF array for all field positions in the scene, an $xy$ meshgrid is defined with a range of $-N_{px}/2$ to $N_{px}/2$. A half pixel shift is then applied to the $xy$ meshgrid to account for the even number of pixels within the scene. The meshgrid is then scaled by the zodiacal scene pixel scale, given by $S_{px}$, such that each entry in the meshgrid now represents the $x$ and $y$ offset of a pixel in the scene. The equations defining the meshgrid creation are given below. 

\begin{equation} \label{eq:scene_meshgrid}
  x = \left(
  \left[ {\begin{array}{ccccc}
  -N_{px}/2 & (-N_{px}+1)/2 & ... & (N_{px}-1)/2 & N_{px}/2\\
  . & . & . & . & .\\
  . & . & . & . & .\\
  . & . & . & . & .\\
  -N_{px}/2 & (-N_{px}+1)/2 & ... & (N_{px}-1)/2 & N_{px}/2
  \end{array} } \right] + 1/2\right) * S_{px}
\end{equation}
\begin{equation}
  y = \left( 
  \left[ {\begin{array}{ccc}
  -N_{px}/2 &  ...  & -N_{px}/2\\
  (-N_{px}+1)/2 &  ...  & (N_{px}-1)/2\\
  . & . & . \\
  (-N_{px}+1)/2 &  ...  & (N_{px}-1)/2\\
  N_{px}/2 &  ...  & N_{px}/2
  \end{array} } \right] + 1/2\right) * S_{px}
\end{equation}

\begin{figure}[H]
    \centering
    \includegraphics[scale=0.55]{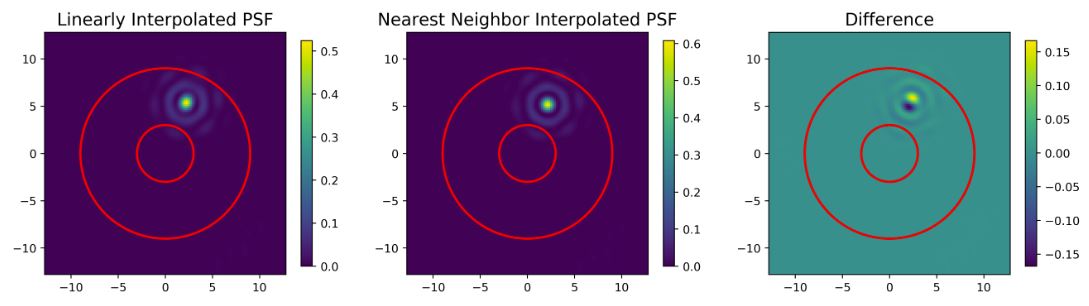}
    \caption{This figure displays the interpolated and rotated PSFs for a source offset of 286.838mas at 66.431\degree with respect to the x-axis along with the difference between the linear and nearest neighbor interpolations.}
    \label{fig:interpped_psfs_example}
\end{figure}

With the $xy$ meshgrid defined in units of milliarcseconds, the final step in creating the interpolated array library is to use the interpolating function setup earlier. For convenience, the $x$ and $y$ meshgrid arrays are flattened such that a single for-loop, which will iterate a total of $N_{px}^2$ times, can be used to generate the PSF array for each pixel offset. In a single iteration of the loop, the $x$ and $y$ offsets are first converted to polar offsets represented by $r$ and $\theta$ using a standard 2D arctangent function. Another meshgrid must also be created which has a range of $0$ to $n_{px}-1$. The $x$ and $y$ components of this meshgrid are then flattened and stacked alongside a vector filled with values of $r$ in units of milliarcseconds. This leaves a set of points with $n_{px}^2$ rows and 3 columns. This set of points is then input into the interpolating function setup earlier, with the output being a flattened grid of the interpolated PSF array for an offset corresponding to the value of $r$. However, this array must be reshaped to a square array with dimensions of $n_{px}$x$n_{px}$. The array is then rotated by the corresponding value of $\theta$. Once rotated, the interpolated PSF array is flattened once again and saved to a data file such that the next iteration within the loop can begin. Flattening this array before saving the data will prove convenient once a simulation is to be done. Once all iterations have been completed, the data file will contain an interpolated PSF for every field position within the zodiacal scene to be simulated. Figure \ref{fig:interpped_psfs_example} shows examples of the interpolated PSFs created using the IPAC PSFs with both linear and nearest neighbor interpolation. These two approaches agree to the $\sim$10\% level. 

\section{Simulating an Image of a Given Scene} \label{sec:simulation}
With the interpolated PSFs created for the given zodiacal scene, image simulations become trivial using the matrix-vector multiplication method. The array of the zodiacal scene must first be flattened and the data of the interpolated PSF array should be loaded into memory if possible. Given matrix-vector multiplication is a linear operation, the assumption of the optical system being linear must be valid. In such a case, multiplying the vector containing the flattened array of the zodiacal scene by the matrix of interpolated PSFs will yield the flattened vector of the image simulation, which can then be reshaped into the correct dimensions. Therefore, it is important to confirm that the process of flattening the array of the zodiacal scene and then reshaping into the correct pixel dimensions does not mistakenly transpose the image or alter the original values of coordinates in any way. In this scenario, the matrix will have $n_{px}^2$ rows and $N_{px}^2$ columns while the flattened zodiacal array will be a single column vector with $N_{px}^2$ entries. Each column in the matrix corresponds to a flattened interpolated PSF for the corresponding $x$ and $y$ offset of the field point in the flattened zodiacal scene. The simulated image will first be a vector with $n_{px}^2$ entries and once reshaped, will be a $n_{px}$ by $n_{px}$ array. The more precise mathematical variables and equations are given below. 

\begin{equation}\label{eq:matrix-vector-descriptions}
    \begin{aligned}
        &\text{Column vector containing a flattened interpolated PSF array:}\ \mathbf{I}_j\ \text{where $j=1,2,...N_{px}^2$}\\
        &\text{Column vector containing a flattened array of a zodiacal scene:}\ \mathbf{Z}\\
        &\text{Matrix containing the interpolated PSF arrays:}\ A\\
        &\text{Flattened simulated image of the zodiacal scene:}\ \mathbf{S}\\
    \end{aligned}
\end{equation}


\begin{equation}\label{eq:interpolated_matrix}
    A = \left[ {\begin{array}{cccccc}
  \mathbf{I}_{1}&\mathbf{I}_{2}&.&.&.&\mathbf{I}_{N_{px}^2}
  \end{array} }
\right]
\end{equation}

\begin{equation}
\mathbf{S} = A\mathbf{Z}
\end{equation}

\begin{figure}[H]
    \centering
    \includegraphics[scale=0.55]{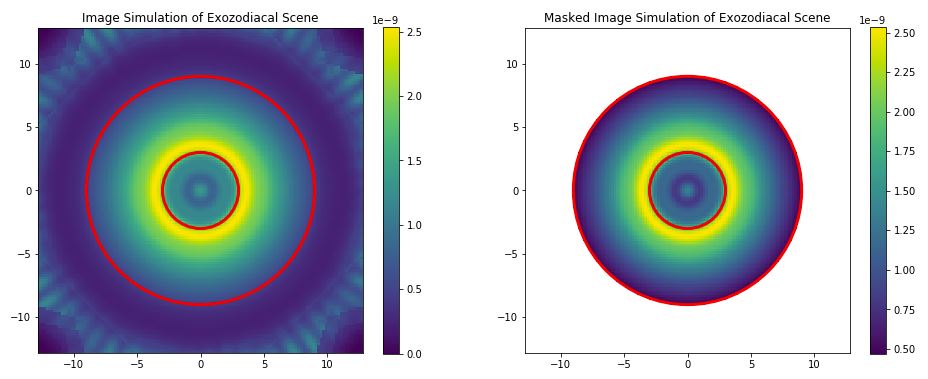}
    \caption{Simulated image of the same zodiacal scene previously shown in Figure \ref{fig:zodi_example}.}
    \label{fig:simulation_example}
\end{figure}

If the computational memory available does not allow for the PSF data to be loaded, then a simple for-loop can also be implemented to read data from the matrix row by row and multiply each row by the value of the corresponding field point of the zodiacal scene. The product can then be added to the existing image vector on each iteration of the loop. While this yields the same numerical result, it will require more time to produce each individual image simulation, however, it forgoes the time required to load the matrix data into memory and allows operation on systems with less available memory than our test system (a single node on the University of Arizona Ocelote High-Performance Computing cluster). An example of a simulated image is shown in Figure \ref{fig:simulation_example}

The units within the final simulation are then derived by multiplying the units of the zodiacal scene, being Janskys, by the units of the PSFs, which may be normalized in various ways. It should be noted this process of image simulation does not take into account any possible detector affects such as saturation or bleeding from pixel to pixel and solely provides the flux information of the final image. 

\begin{figure}[h]
    \centering
    \includegraphics[scale=0.45]{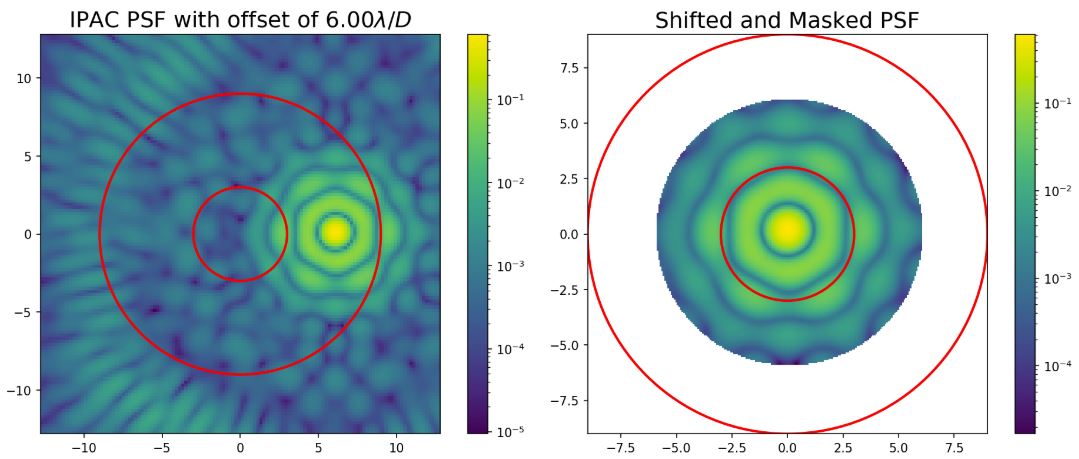}
    \caption{PSF for a source offset of $6\lambda/D$ which is then masked and shifted to be used for convolution with the zodiacal scene.}
    \label{fig:convolution PSF}
\end{figure}

In addition to the simulation shown above, the more standard method of simulation by convolution is also preformed using the convolve$\_$fft() function providied by Astropy. In this case, a PSF with a source offset of $6\lambda/D$ (the center of the dark-hole region) is masked and shifted to act as the PSF of the center of the FOV as shown in Figure \ref{fig:convolution PSF} in order to be used for convolution.

While this method of simulation can be performed very quickly, approximately in 0.05s according to the timing on the computational node being used, this method requires one additional assumption causing it to lack accuracy. The assumption is that no flux from within the inner working angle contributes to the flux at the final image.

\begin{figure}[h]
    \centering
    \includegraphics[scale=0.55]{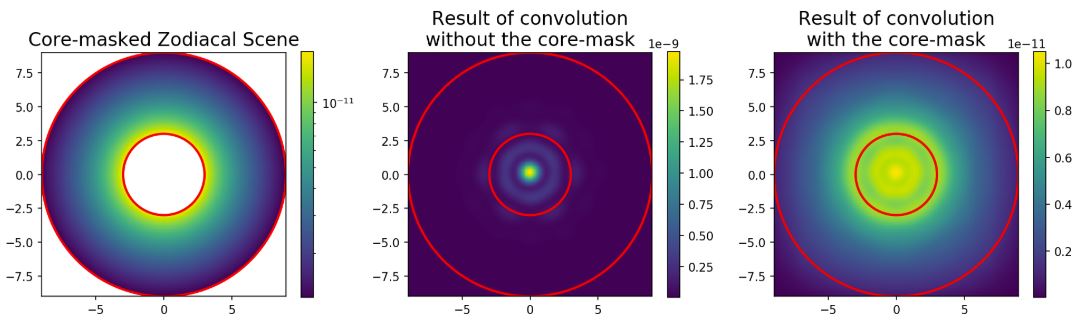}
    \caption{The left array is that of the same zodiacal scene which was shown before, except with a core-mask applied to it such that when a simulation is run, no flux from within the inner working angle would be included. The middle and right arrays are simulations performed by convolution displaying how the result looks very similar to the PSF itself if the core is not masked due to the brightness of the central portion of the scene. If the core is masked, then the structure within the image is more similar to that which is expected, however, more light still leaks inside the inner working because of the outer rings of the PSF used for convolution.}
    \label{fig:coremasked_sim}
\end{figure}

Figure \ref{fig:coremasked_sim} displays the results of convolution of a scene with an off-axis PSF. From the figure, it is clear that because the center of the model scene is much brighter than the outer regions of the scene, which are of greater scientific interest, the convolution yields a simulation looking very similar to the PSF itself. Therefore, in order to obtain a simulation with a much more similar shape to that which is expected, the center of the zodiacal scene must be masked. The benefit of using the interpolated arrays is precisely that the field-dependent nature of the PSFs is taken into account, meaning the flux of from within the inner-working angle does contribute to the final image, making it more accurate to the performance of the real system. 

\section{Results}\label{sec:results}
The results discussed will include the time required to generate the interpolated PSF grid followed by the time required to generate the image simulations using the interpolated PSFs with matrix-vector multiplication as well as a standard for-loop. The simulations using the interpolated PSF grids are then compared to the simulations which do not use the PSF grids. As computational capabilities vary across all computers, the specifications for the computational configuration used on a University of Arizona Ocelote HPC node are given below.

\begin{table}[H]
\caption{Specifications for the computational configuration used for the generation of the interpolated PSFs and the simulated images.}
    \centering
    \begin{tabular}{|c|c|c|c|} \hline
    Model & Number of CPU Cores & Processor Speed & Memory \\ \hline
    Lenovo NeXtScale nx360 M5 & 8 & 2.3GHz & 192GB  \\ \hline
    \end{tabular}
    \label{tab:my_label}
\end{table}

\subsection{Results for Creating the Interpolated PSF Grids} \label{sec:results_creation}
Multiple sets of PSFs were used to create the simulations of the zodiacal scenes. One of these PSF sets, which was described above and available on the \href{https://roman.ipac.caltech.edu/sims/off_axis_PSF.html}{Roman IPAC website}, used pre-phase-B pupil, phase-A OTA misalignment, OTA+CGI surface aberrations, and phase-B polarization effects. A full Fresnel diffraction model along with a wavefront control system was implemented to produce these PSFs. Another set of PSFs was created with the use of the PROPER module for python along with the wfirst$\_$phaseb$\_$proper\footnote{Python Version 1.7, https://github.com/ajeldorado/proper-models/tree/master/wfirst$\_$cgi/models$\_$phaseb/python} module created by John Krist, which uses solely phase-B pupil and FPM data. These PSFs are 256 by 256 grids with a pixel scale of 0.1$\lambda/D$ such that the total extent is the same as the PSFs from IPAC. The source offsets for these PSFs were made more dense such that interpolation, whether it be linear or nearest, would also be more accurate. The source offsets chosen ranged from $0\lambda/D$ to $11\lambda/D$ with an equispaced sampling of $0.05\lambda/D$, although much of the flux of PSFs beyond $9\lambda/D$ is lost due to the fact a feild stop is implemented to reduce the FOV to the outer working angle. This set of PSFs used optical phase errors but no deformable mirror maps or actuated pistons. Lastly, another set of PSFs with the same dimensions, pixel sampling, source sampling, and optical phase errors as that of the aforementioned were created, but this time with the deformable mirror pistons activated. In both set of PSFs created using the PROPER module, the mean of all polarization mode aberrations was implemented. 

The table below displays the information regarding the creation of the interpolated PSF arrays using each set of PSFs for both linear and nearest interpolation. The results are found by creating the interpolated arrays 10 times and taking the average of the results. The results are consistent with nearest neighbor interpolation requiring approximately 2/3 of the time required for that of linear interpolation. The time required for interpolation of the PSFs created with the PROPER module is also consistent in that it is greater than 4 times that of the time required for the IPAC PSFs no matter which method of interpolation was used. Given the PROPER PSFs contain twice the amount of pixels on each axis as those from IPAC, this means the time required for the generation is on the order of $n_{px}^2$.

\begin{table}[H]
\caption{Timing tests for the creation of the interpolated PSFs along with file size of the interpolated arrays.}
    \centering
    \begin{tabular}{|P{5cm}|P{2cm}|P{2cm}|P{2cm}|P{1.5cm}|} \hline
    PSFs Used & Array Dimensions & Interpolation Method & Time to Create [s] & Size of File [GB] \\ \hline
    IPAC & 128x128 & Linear & 506.680 & 8 \\ \hline
    IPAC & 128x128 & Nearest & 334.246 & 8 \\ \hline
    PROPER & 256x256 & Linear & 2376.893 & 32 \\ \hline
    PROPER & 256x256 & Nearest & 1507.422 & 32 \\ \hline
    PROPER with DM maps & 256x256 & Linear & 2337.210 & 32 \\ \hline
    PROPER with DM maps & 256x256 & Nearest & 1458.876 & 32 \\ \hline
    \end{tabular}
    \label{tab:results_interpped_table}
\end{table}

For context, the PSF data cubes created with the PROPER module using propagation techniques required an average of 4287s to create. Given the $0.05\lambda/D$ offset sampling of those PSFs, there were a total of 221 PSFs created for each set of PROPER PSFs. Therefore, if propagation were to be used to create a set of PSFs for every field-point in the zodiacal scenes, the required amount of time would be 1,271,280s or 353.13hrs given there are a total of $N_{px}^2$ field-points in the scenes. This demonstrates the benefit of using interpolation and rotation of previously propagated PSFs to create all necessary PSFs for simulation. 

\begin{figure}[h]
    \centering
    \includegraphics[scale=0.55]{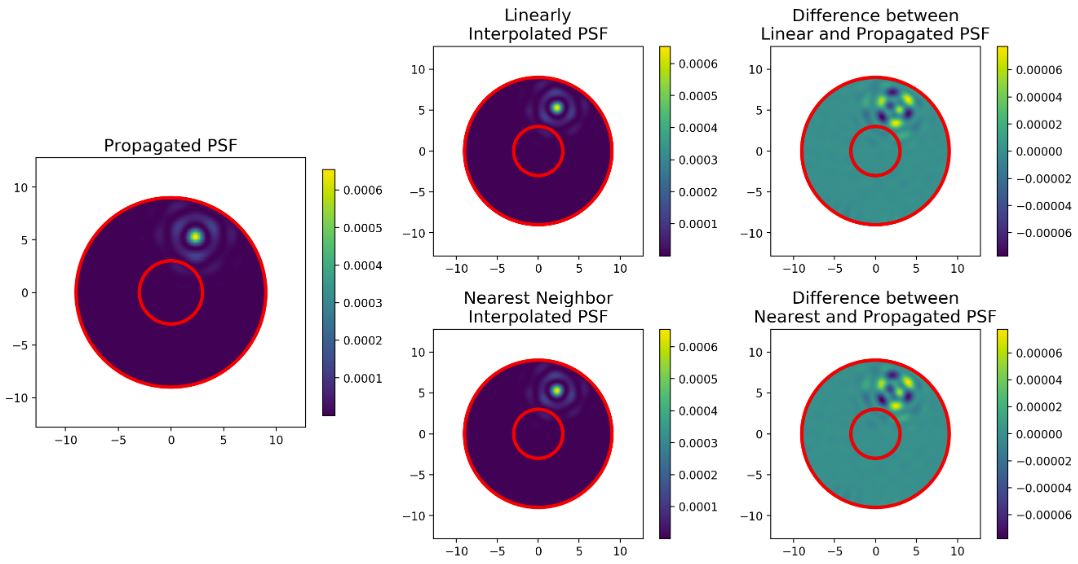}
    \caption{PSF created from propagation for a source offset of 286.838mas at 66.431\degree with respect to the x-axis is compared to the PSFs created from interpolation and rotation to simulate a source for such an offset.}
    \label{fig:interpped_psf_vs_propagated_psf}
\end{figure}

In addition, Figure \ref{fig:interpped_psf_vs_propagated_psf} shows a comparison of a PSF with both an $x$ and $y$ source offset created via propagation from the PROPER module versus both the linear and nearest interpolations done to create the PSF for the same source offset. From the differences, it is clear that there is no significant difference between the linear and nearest interpolations given the similarity between the difference arrays. Rather, the largest differences are due to the rotation of the PSF not properly accounting for the diffraction effects which cause the bright areas of the outer rings to rotate differently. 

\subsection{Results for the Simulations} \label{sec:results_sims}
When comparing the results of the simulations, only data within 9$\lambda/D$ is compared as the rest of the data is outside of the dark hole region of the HLC, so it is not as significant to the scientific purpose of the mission. Table \ref{tab:load_interpped_table} displays the time required to load the matrix data into memory given the memory capacity of the computer is adequate. From this, it is consistent that the time required to load that data increases by $n_{px}^2$, similar to the time required to create the matrix data. Table \ref{tab:results_sims_table} displays the averages of time required to compute the simulations. It is clear that if previously computed interpolated PSFs are not used, then the time required for a simulation is approximately the same amount of time as the creation of the interpolated PSFs matrix. Therefore, this presents the issue for running many simulations. However, if a simple for-loop is implemented to perform the simulation using the previously computed interpolated PSFs, then the time required is on the order of 10s depending on the array size of the PSFs being used. Even quicker to run simulations is to use the matrix data by previously loading it into memory, allowing for simulations to be done in under 1s. 

\begin{table}[H]
    \centering
    \caption{Timed results for loading the matrix data of the interpolated PSFs into memory.}
    \begin{tabular}{|P{6cm}|P{3cm}|} \hline 
    Interpolated PSFs Used & Time Required to load Matrix Data into Memory [s] \\ \hline
    IPAC, Linear & 26.63 \\ \hline
    IPAC, Nearest & 26.60 \\ \hline
    PROPER, Linear & 118.44 \\ \hline
    PROPER, Nearest & 121.11 \\ \hline
    PROPER with DM maps, Linear & 112.95 \\ \hline
    PROPER with DM maps, Nearest & 125.85 \\ \hline
    \end{tabular}
    \label{tab:load_interpped_table}
\end{table}

\begin{table}[H]
    \centering
    \caption{Timed results for simulating the images using the various interpolated PSF arrays as well as without using the arrays. All values are in [s].}
    \begin{tabular}{|P{4cm}|P{3.5cm}|P{3.5cm}|P{4cm}|} \hline 
    Interpolated PSFs Used & Time Required for simulation using Matrix-Multiplication & Time required using a for-loop & Time Required for simulations using Standard Interpolation Method \\ \hline
    IPAC, Linear & 0.1642 & 5.858 & 496.2106\\ \hline
    IPAC, Nearest & 0.1651 & 6.517 & 339.053 \\ \hline
    PROPER, Linear & 0.5156 & 18.950 & 2413.792 \\ \hline
    PROPER, Nearest & 0.5138 & 17.948 & 1403.009\\ \hline
    PROPER with DM maps, Linear & 0.4977 & 17.970 & 2222.512\\ \hline
    PROPER with DM maps, Nearest & 0.5414 & 16.993 & 1374.960\\ \hline
    \end{tabular}
    \label{tab:results_sims_table}
\end{table}

\begin{figure}[h]
    \centering
    \includegraphics[scale=0.5]{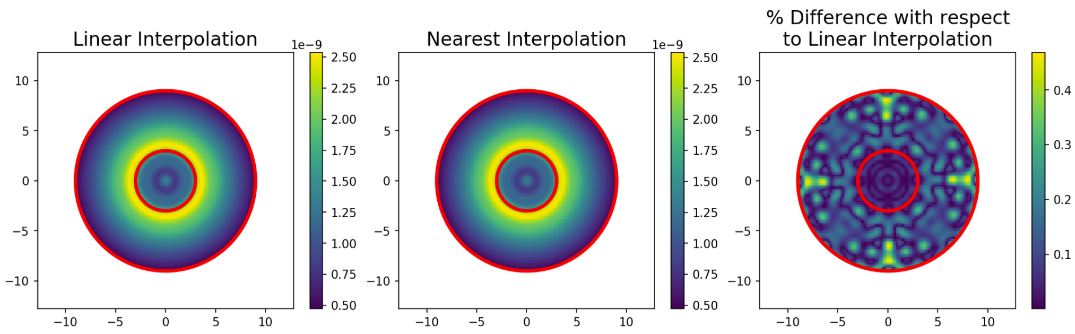}
    \caption{Results of the simulations done using the publicly available IPAC PSFs along with the $\%$ difference between results.}
    \label{fig:ipac_sims_difference}
\end{figure}

Below are figures showing comparisons between results of the simulations when using the different interpolation methods. Figure \ref{fig:ipac_sims_difference} shows the difference between simulations created using the linear and nearest neighbor interpolation data of the IPAC PSFs. The results show a maximum percent difference of $0.468\%$ and an average percent difference of $0.114\%$. 

\begin{figure}[H]
    \centering
    \includegraphics[scale=0.5]{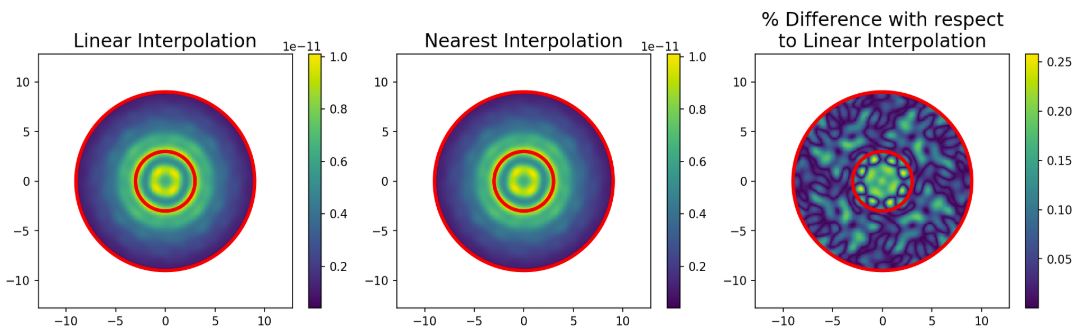}
    \caption{Results of the simulations done using the PROPER PSFs generated without DM wavefront maps along with the $\%$ difference between results.}
    \label{fig:prop_sims_difference}
\end{figure}

\begin{figure}[H]
    \centering
    \includegraphics[scale=0.5]{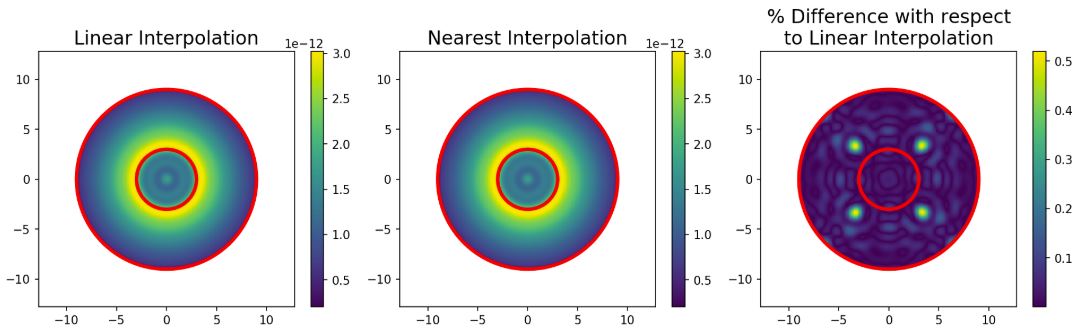}
    \caption{Results of the simulations done using the PROPER PSFs generated with DM wavefront maps along with the $\%$ difference between results.}
    \label{fig:propDMs_sims_difference}
\end{figure}

Figure \ref{fig:prop_sims_difference} shows similar results, except using the simulations from the PROPER PSFs which did not employ any deformable mirrors. The simulation results themselves display the importance of the deformable mirrors and wavefront control for the optical system as the image has much more flux within the inner working angle due to the phase errors present in the optics. However, the maximum percent difference between the simulations using linear and nearest interpolation is $0.258\%$ with an average percent difference of $0.0598\%$. Very similar values are found when performing the simulations using the PROPER PSFs which had activated DMs, which is shown in Figure \ref{fig:propDMs_sims_difference}. In these simulation results, the overall flux pattern is much more similar to that of the IPAC PSFs; however, there is a scaling difference between the PSF normalization causing the flux with the IPAC PSFs to appear about 1000x brighter.


For some of the simulations done using the PROPER PSFs, the maximum percent difference would be higher than that of the IPAC PSF simulations. However, the average percent difference between the linear and nearest simulations is about 1/3 of the simulations performed with IPAC PSFs. This is likely a result of the finer sampling used when creating the PROPER PSFs, resulting in the linear and nearest interpolations to be more alike than the linear and nearest interpolations of the IPAC PSFs. While outside the inner working angle the PSF shape does not evolve much, the flux of the PSF can have a significant variation between source offsets. Overall, however, the difference between the interpolation methods for the PSFs does not seem to make much of a difference when a simulation is created as within all cases, there is less than a $1\%$ difference within the arrays.

\begin{figure}[H]
    \centering
    \includegraphics[scale=0.3]{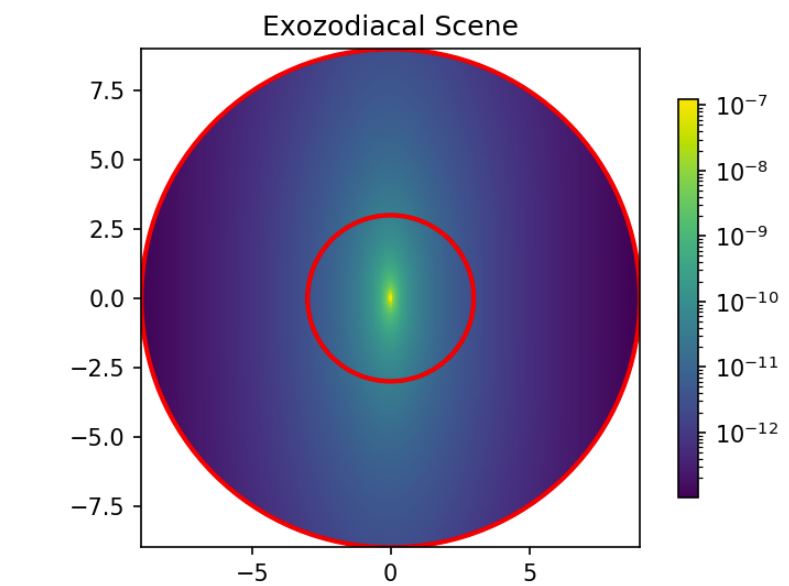}
    \caption{This figure displays an example of another zodiacal scene generated with the same parameters as before, except with an inclination of 81.29\degree.}
    \label{fig:zodi_example_2}
\end{figure}

In order to confirm whether or not the the performance of the interpolated PSFs is consistent with other dust models, the matrix-vector multiplication method was applied to generate simulations for models with various inclinations. Figure \ref{fig:zodi_example_2} displays another example of a zodiacal scene with the same parameters as the one shown before, only with an inclination of 81.29\degree. The figures below display similar results as far as the simulations using the different interpolation methods as well as the $\%$ difference between the simulations. The results remained consistent with the maximum percent difference between the linear and nearest simulations remaining below $1\%$ and the average of the percent difference when using the PROPER PSFs remaining approximately 1/3 of the simulations using IPAC PSFs. 

\begin{figure}[H]
    \centering
    \includegraphics[scale=0.5]{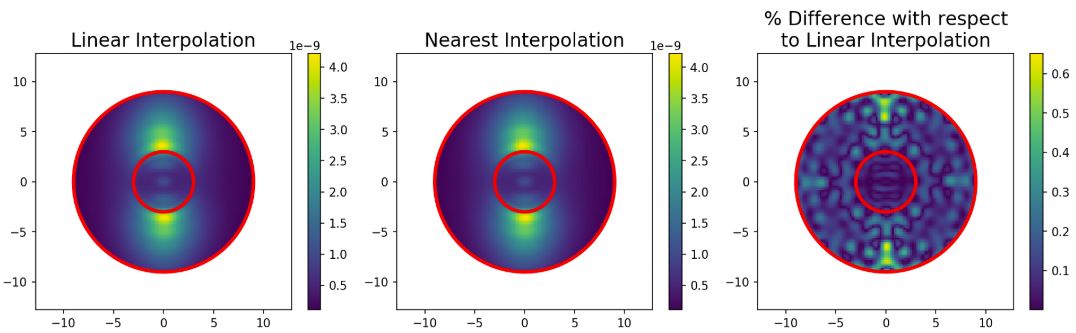}
    \caption{Results of the simulations of the scene with an inclination of 81.29\degree done with the IPAC PSFs.}
    \label{fig:ipac_sims_difference_2}
\end{figure}

\begin{figure}[H]
    \centering
    \includegraphics[scale=0.5]{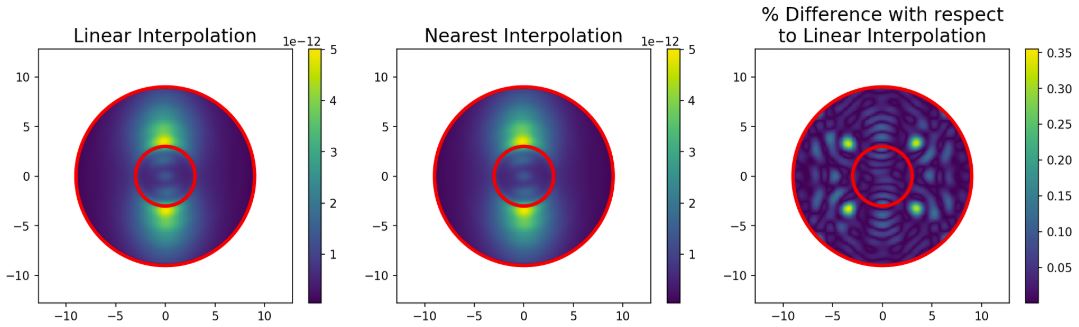}
    \caption{Results of the simulations of the scene with an inclination of 81.29\degree done with PROPER PSFs with DM maps.}
    \label{fig:prop_sims_difference_2}
\end{figure}

\section{Conclusions}
This work describes an efficient approach to generating images of extended sources that captures the field-dependent effects of a coronagraph PSF. 
On a system with sufficient memory, we find direct matrix multiplication with interpolated and rotated PSFs to be more accurate than a convolution with an off-axis PSF within the working angles of the system. In addition, by using previously computed interpolated PSF arrays to perform the matrix multiplication, the time required to generate the simulation is on the same order as the time required for a simulation done with convolution. Furthermore, using interpolation to create the necessary PSFs for simulating a scene can generate the PSFs hundreds of times faster than using a complete diffraction model. 

We also find generation of the matrix operator can be performed efficiently via nearest neighbor interpolation, given a sufficiently sampled grid, regardless of input disk geometry. This method allows for simulations of a variety of models to be completed efficiently, proving to be important in applications such as simulating multiple complex astrophysical scenes at once or for accounting for instrumental effects when fitting parameters of an exozodiacal scene from future Roman coronagraphic images. Future work will add noise to debris disk images and demonstrate retrieval of exozodiacal disk properties.

\section{Acknowledgements}
The authors acknowledge valuable inputs from  Vanessa Bailey, John Debes, Brian Kern, John Krist,  Bertrand Mennesson,  Hanying Zhou,  and the rest of the JPL and IPAC CGI teams. 
We also express gratitude to Jaren Ashcraft and Bianca Alondra Payan for helpful discussions. 

Portions of this work were supported by the WFIRST Science Investigation team prime award $\#$NNG16PJ24C.
Portions of this work were supported by the Arizona Board of Regents Technology Research
Initiative Fund (TRIF).
This research made use of the High Performance Computing (HPC) resources supported by the University of Arizona (UA) TRIF, UITS, and RDI and maintained by the UA Research Technologies department.

This research made use of community-developed core Python packages, including: Astropy \cite{the_astropy_collaboration_astropy_2013}, Matplotlib \cite{hunter_matplotlib_2007}, SciPy \cite{jones_scipy_2001}, Jupyter and
the IPython Interactive Computing architecture \cite{perez_ipython_2007,kluyver_jupyter_2016}.
\bibliographystyle{spiebib} 
\bibliography{mybib,wfirst} 
\end{document}